\documentclass[a4paper]{jpconf}
\usepackage{graphicx}
\begin{document}
\title{Measurement of Longitudinal Spin Asymmetries From $W\rightarrow e$ Boson Decay in Polarized pp Collisions at $\sqrt{s}=500$ GeV at RHIC-PHENIX}

\author{Kensuke Okada (for the PHENIX collaboration)}

\address{BNL Bldg.510A, Upton, NY 11973, USA}

\ead{okada@bnl.gov}
% 5 pages for a regular talk
% /phenix/WWW/p/info/dp1/120/1resub

\begin{abstract}
We report the measurement of the parity violating single spin 
asymmetries for inclusive high transverse momentum electrons and 
positrons in polarized $p+p$ collisions at a center of mass energy 
of $\sqrt{s}=500$ GeV with the PHENIX detector at RHIC. 
These electrons are attributed to the decay of $W^\pm$ and $Z^0$ bosons,
and measured production cross section is consistent with the 
expectations. The $W$ production is confirmed for the first time 
in $p+p$ collisions. Its spin asymmetry in the polarized $p+p$ 
collisions is a 
important probe for the quark flavor decomposition of the proton spin.
\end{abstract}

% Introduction
 The proton is the most basic composite particle. Understanding 
 its structure helps us gain insight into the quark 
confinement. Deeply inelastic scattering (DIS) experiment has been 
a clear and powerful approach and it is evolved to polarized DIS 
experiment to explore the spin structure of the proton. 
Analyses of polarized semi-inclusive DIS
experiments~\cite{Alekseev:2010ub,Airapetian:2004zf,Adeva:1997qz}
have determined the individual flavor separated quark and antiquark 
helicity distribution ($\Delta q$ and
$\Delta \bar{q}$) by connecting final state hadrons with quark
flavors using fragmentation functions.
Colliding polarized protons is a complementary way to approach 
the origin of the proton spin. At the collider energy, the real 
$W$ is produced via a parity violating weak process, which enables 
to identify the quark flavor and helicity in the proton contributed 
to the process
by detecting decay leptons without the uncertainty of fragmentation 
functions. Another advantage is, because the scale is set by the heavy 
mass of the $W$, higher order QCD corrections can be evaluated 
reliably.

 In 2009, the relativistic heavy ion collider (RHIC) succeeded to 
collide polarized protons at a center of mass energy of 
$\sqrt{s}=500$ GeV. In this report,
we describe the measurement of the production cross section of 
$W$ boson and its parity violating 
single spin asymmetry.

% Run9 and PHENIX detector 
The PHENIX detector has been described in detail elsewhere~\cite{Adcox:2003zm}.
The central arm spectrometer covers $\left|\eta\right|<0.35$ in pseudorapidity.  
This analysis uses the electromagnetic calorimeter (EMCal) to 
measure the energy of electrons and tracking chambers, the drift 
chambers (DC) and the pad chambers (PC), to determine the charge sign 
of the tracks from their bend angle in an axial magnetic field. 
The data set was recorded with the EMCal trigger which has a nominal 
energy threshold at 10 GeV. 
The luminosity is monitored by beam-beam counters, those are 
two arrays of 64 quartz \v{C}erenkov counters located at 
3.1$<\left|\eta\right|<$3.9. Their coincidence rate is connected 
to the luminosity from the van der Meer scan technique~\cite{Adare:2008qb}.
In the 2009 run, the integrated 
luminosity of 8.6 pb$^{-1}$ was recorded for the central arm detector analysis.

\begin{figure}[h]
\begin{minipage}{18pc}
\includegraphics[width=18pc]{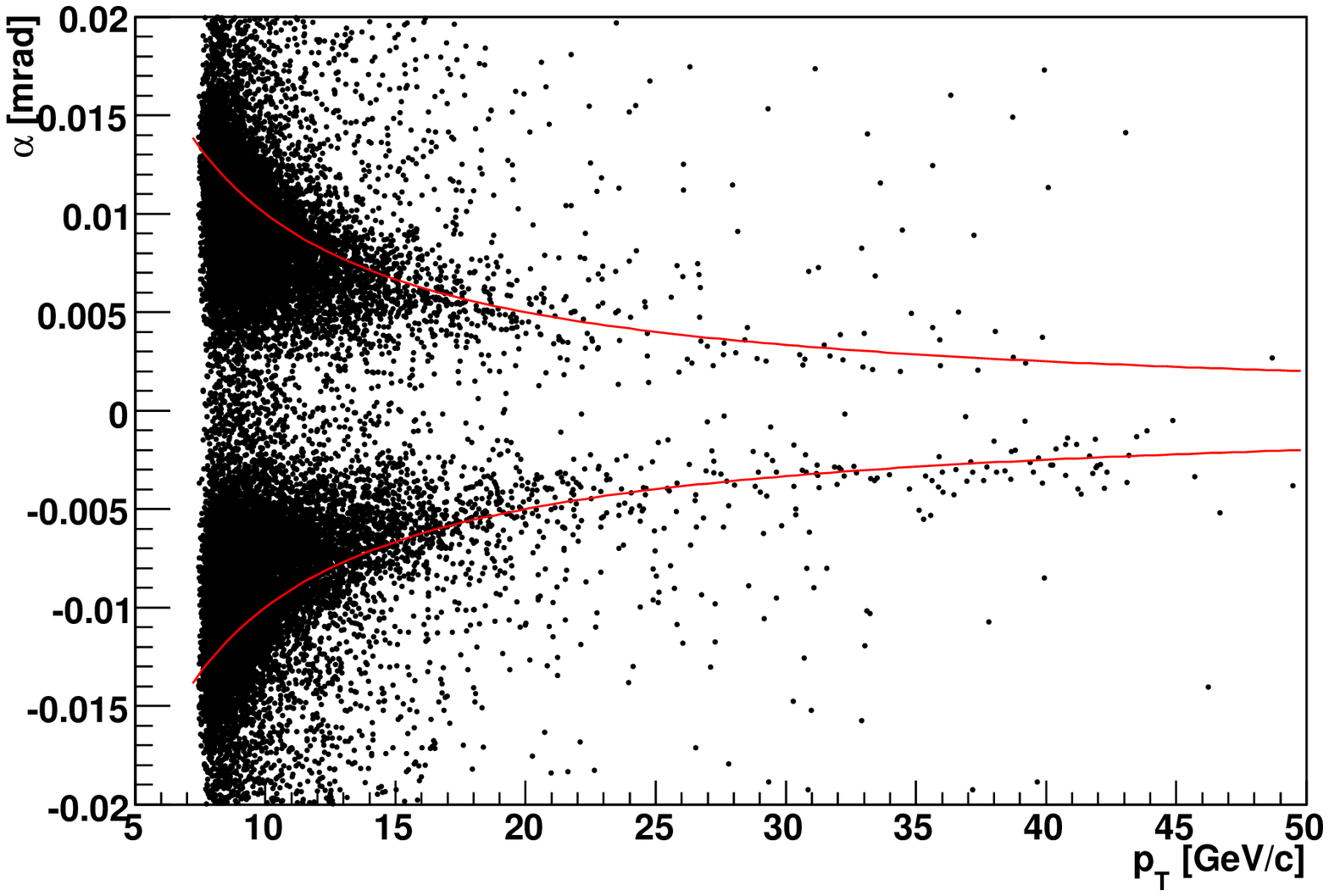}
\caption{\label{fig:alphapt}
The correlation between the bend angle ($\alpha$) and the transverse 
momentum ($p_t$). The lines show the expectation.}
\end{minipage}\hspace{2pc}%
\begin{minipage}{18pc}
\includegraphics[width=18pc]{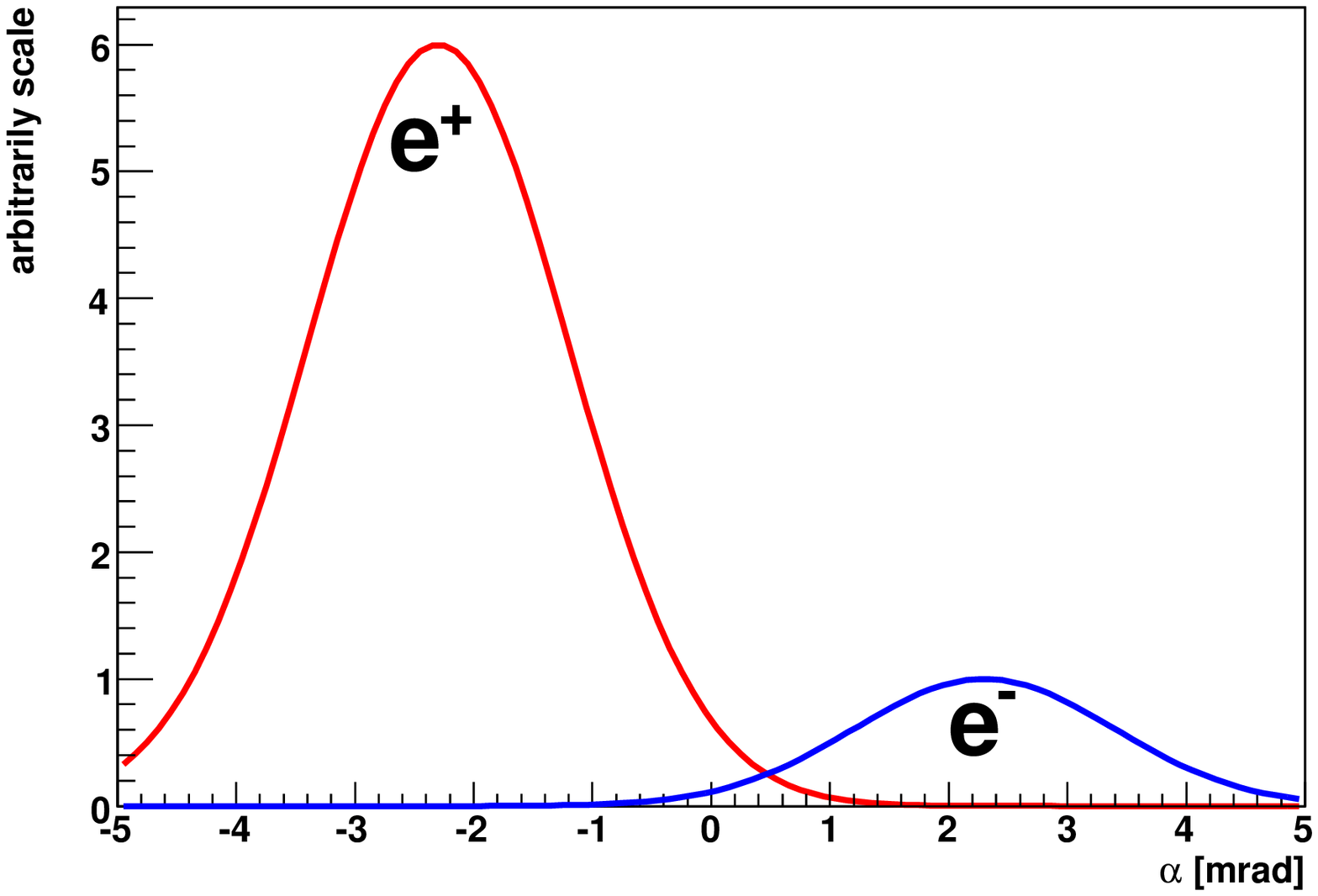}
\caption{\label{fig:alpha40gev} The $\alpha$ distribution for 40GeV track. The ratio of two Gaussian distribution is set to the expected signal ratio of $W^+$ and $W^-$. }
\end{minipage} 
\end{figure}

% Analysis
This analysis searched for electrons from the $W$ decay. 
The data collected by the EMCal trigger are mostly QCD events. 
In the $W$ event, the electron has high transverse momentum ($p_T$) 
and it is isolated unlike high $p_T$ track in QCD jet events.
The electron energy measured in the EMCal was used to calculate 
the $p_T$. 
Because for electrons at momentum of 40GeV/$c$ of our interest, 
the resolution of the energy in the EMCal is about 5\% and it is 
better than the resolution of the momentum of about 40\%. 
Figure \ref{fig:alphapt} shows the relation 
between the bend angle, $\alpha$, and the transverse momentum
for charged tracks after a rough electron selection. Though there are 
contaminations of hadron backgrounds in the low $p_T$ region, the 
data points are seen to be distributed around the 
calculation from the magnetic field strength. 
The resolution of the tracking system was evaluated from the data with 
no magnetic field. Figure \ref{fig:alpha40gev} shows the $\alpha$ 
distribution for 40 GeV/$c$ tracks. It is capable to determine 
the charge sign reasonably.
The data with no magnetic field were also used to determine the 
center of transverse beam position.

% Spectra Jacobian peak
To select electron track candidates we applied two levels of criteria. 
The first one is a minimal track criteria which requires the projected 
point to the beam to be less than 30 cm, loose cuts on the time of 
flight, and rough energy-momentum matching. This criteria was used to 
calculate the production cross section.
The second one requires the electron track to be isolated. 
The threshold was set to the sum of energy and momentum in the cone 
around the electron candidate track at 2 GeV. 
This cut purifies the signal fraction. However it is hard to evaluate 
the cut efficiency, so this sample was only used to calculate the spin 
asymmetry.

\begin{figure}[h]
\includegraphics[width=20pc]{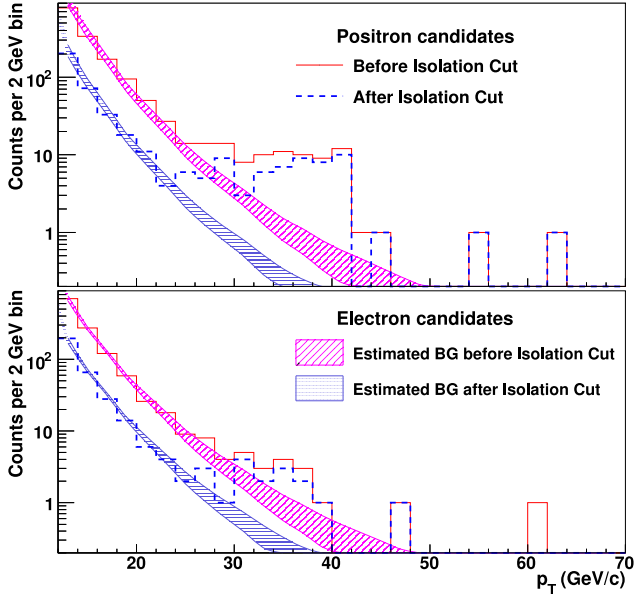}\hspace{2pc}%
\begin{minipage}[b]{14pc} \caption{\label{fig:espectra}(color online) The spectra of positron (upper panel)
and electron (lower panel) candidates before (solid histogram)
and after (dashed histogram) an isolation cut. The estimated
background bands are also shown.
}
\end{minipage}
\end{figure}

Figure \ref{fig:espectra} shows the $p_T$ spectra for positron and 
electron candidates. The bands represent our estimated background, 
which are dominated by charged hadrons with hadronic interactions 
in the EMCal 
and electrons from photon conversions before the tracking system.
In the figure, clear electron signals from the $W$ decay are seen 
by the Jacobian peak at $M_W/2 \simeq 40$ GeV.
It has to be noted that our selection is limited to one hemisphere, 
so that the signal is a combination of $W$ and $Z$ bosons decay.

% Cross section
 From the yields in the signal region ($30\!<\!p_T\!<\!50$ GeV/$c$), 
the production cross sections were calculated. 
   %%  from ppg120
The tracks within the nominal geometric acceptance of the 
central spectrometer were reconstructed with $\sim$37\% 
efficiency defined by the overlap of live areas in the 
tracking detectors, and fiducial areas on the calorimeters and 
drift chambers. The efficiency for retaining electron 
candidates after all cuts was 99\%. The resulting 
reconstruction efficiency was not $p_T$ dependent for 
$p_T\!>\!30$ GeV/$c$.
Table \ref{table:xsecresult} shows the production cross 
sections for electrons and positrons from $W$ and $Z$ boson 
decays. 
The first error is statistical; the second error is 
systematic from the uncertainty in the background; 
and the third error is a normalization uncertainty.  
The normalization uncertainty is due to the luminosity (10\%),
multiple collision (5\%), and acceptance and efficiency
uncertainties (10\%).
The results are compared with NLO and NNLO calculations.
The difference of the two calculations indicates a level of 
theory uncertainties.
Within the uncertainty, our measurements are consistent with 
expectations. From theory calculations, the fraction of $Z$ 
boson contributions to our sample are 7\% and 31\% for positrons 
and electrons, respectively. 

\begin{center}
\begin{table}[h]
\caption{\label{table:xsecresult} 
Comparison of measured cross sections for electrons 
and positrons with $30<p_T<50$ GeV/$c$ from $W$ and $Z$ 
decays with NLO~\cite{deFlorian:2010aa,Nadolsky:2003ga} 
and NNLO~\cite{Melnikov:2006kv} calculations. 
The first error is statistical; the second error is 
systematic from the uncertainty in the background; 
and the third error is a normalization uncertainty.  
}
\centering
\begin{tabular}{l c c c} 
\br
 & \multicolumn{3}{c}{$\frac{d\sigma}{dy}(30\!<\!p_{T}^e\!<\!50\,{\rm GeV}/c)|_{y=0}$ [pb]} \\
 Lepton & Data & NLO & NNLO\\
 \mr
 $e^+$ & $50.2 \pm7.2 ^{+1.2}_{-3.6} \pm 7.5$ & 43.2 & 46.8\\
 $e^-$ & $9.7 \pm3.7 ^{+2.1}_{-2.5} \pm 1.5$ & 11.3 & 13.5 \\
 $e^+$ and $e^-$ & $59.9 \pm8.1 ^{+3.1}_{-6.0} \pm 9.0$ & 54.5 & 60.3\\
 \br
 \end{tabular}
 \end{table} 
 \end{center}

% Asymmetry 
To calculate the spin asymmetry, the sample with 
the isolation cut was used to minimize the background contamination. 
To reduce the ambiguity of charge misidentification to a negligible 
level, a further cut was applied to the bend angle ($\alpha$)
to be $|\alpha|>1$ mrad.
When a polarized beam collides with a unpolarized beam, 
the raw parity violating single spin asymmetry is defined by
\begin{equation}
\epsilon_{L} = \frac{N^{+} - R \cdot N^{-}}{N^{+} + R \cdot N^{-}}
\end{equation}
where $N^{+}$ is the number of events from collisions with the beam
polarization is positive, and $N^{-}$ for the negative. 
Generally the integrated luminosities are not equal in the two cases,
the relative luminosity, $R$, is to adjust the difference.  
The physics asymmetry is calculated from the raw asymmetry according to
\begin{equation} 
A_{L} = \frac{\epsilon_{L} \cdot D}{P},
\end{equation} 
where $P$ is the beam polarization and $D$ is a dilution 
correction to account for the remaining background in the 
signal region.

In reality, RHIC has two polarized beams. Since the measurement in 
the central arm spectrometer is symmetric to the beams, both beams 
contribute equally. We built a likelihood function from yields 
sorted by four helicity states.
The statistical uncertainty of the raw asymmetry is 
confirmed to 
be consistent with what we expect from Poisson distribution of two 
times of number of candidates. 

The second column in Table \ref{table:spinasym} shows the measured 
raw asymmetries. For the sample in the background region ($12\!<\!p_T\!<\!20$ GeV/$c$),
the asymmetry is consistent with zero, which is expected from the 
fact that they are dominated by the QCD process. For the signal region
($30\!<\!p_T\!<\!50$ GeV/$c$), large asymmetries were observed. Especially it is 
significant for the positrons.
 
To get the physics asymmetries,
the dilution correction of $D=1.04\pm0.03$ and $1.14\pm0.10$ for 
positive and negative charges, respectively, and the average beam 
polarization $P=0.39\pm0.01$ were applied. 
The longitudinal 
polarization fractions were monitored using very forward 
neutron asymmetries~\cite{Adare:2007dg} and found to be 99\% 
or greater. The contribution to $A_L$ from the residual 
transverse component of the polarization was negligible thanks to the 
almost left-right symmetric detectors. 
Table \ref{table:spinasym} shows the results of $A_L$ and its confidence
intervals, those contains the effect of broadening of the likelihood function due to the uncertainties of $D$ and $P$.
When the confidence 
interval was calculated, the physical boundary ($A_L=\pm1$) was 
applied.  With limited statistics, one side of 68\% CL and 95\% CL 
hits this physical boundary. 
A non-zero parity violating single spin asymmetry is observed 
in positrons.

\begin{center}
\begin{table}[h]
\caption{
Longitudinal single-spin asymmetries.  The confidence intervals
are defined for $A_{L}$.
\label{table:spinasym}}
\centering
\begin{tabular}{l c c c c}
\br
Sample & $\epsilon_{L}$ & $A_L(W\!\!+\!\!Z)$ & 68\%\,CL & 95\%\,CL \\
\mr
Background + & $-0.015 \pm 0.04$ &  &  &  \\
Signal + & $-0.31 \pm 0.10$ & $-$0.86 & [$-1,-0.56$] & [$-1,-0.16$] \\
\mr
Background $-$ & $-0.025 \pm 0.04$ &  &  &  \\
Signal $-$ & $0.29 \pm 0.20$ & +0.88 & [$0.17,1$] & [$-0.60,1$] \\
\br
\end{tabular}
\end{table}
\end{center}

\begin{figure}[h]
\includegraphics[width=24pc]{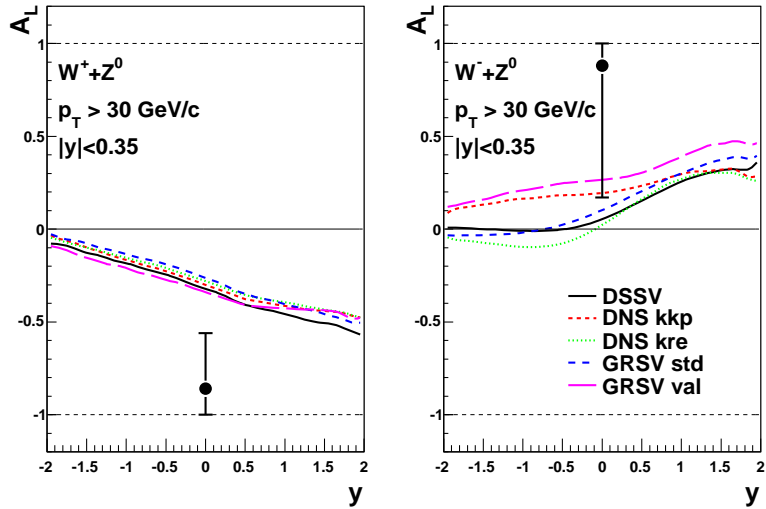}\hspace{2pc}%
\begin{minipage}[b]{10pc}\caption{\label{fig:alfig}(color online) 
Longitudinal single-spin asymmetries for electrons and  
positrons from $W$ and $Z$ decays. The error bars represent 
68\% CL. The theoretical curves are calculated using NLO with 
different polarized PDFs.
}
\end{minipage}
\end{figure}

% Discussion and summary
The results are compared with estimations of various polarized 
parton-distribution functions (PDFs)~\cite{deFlorian:2010aa} in 
Fig. \ref{fig:alfig}. With the current limited statistics, 
the measured asymmetries are consitent with estimations.
Another message from this figure is that the 
admixture of $W$ and $Z$ bosons can be a probe to separate models.

In summary, we presented the first measurement of production 
cross section and non-zero parity violating asymmetry in $W$ and 
$Z$ production in polarized $p+p$ collisions at $\sqrt{s}=500$ GeV.
A non-zero spin asymmetry in positron candidates is a direct 
demonstration of the prity-violating coupling of the $W$ to the 
light quraks.
In the following years, a precise measurement 
is the main goal in the RHIC spin program. PHENIX
is preparing a detector upgrade which enables us to measure the 
forward and backward muons from the $W$ decay. 
We will get more insight into
flavor separated quark and antiquark helicity distributions
in the proton.

\section*{References}

%\begin{thebibliography}{9}
%\bibitem[Adcox et~al.(2003)]{Adcox:2003zm}
%\bibinfo{author}{K.~Adcox} \it{et~al.},
%\bibinfo{journal}{Nucl. Instrum. Meth.} \textbf{\bibinfo{volume}{A499}},
%\bibinfo{pages}{469} (\bibinfo{year}{2003}).
%\end{thebibliography}

%\bibliographystyle{plain}
\bibliographystyle{iopart-num}
%\usepackage{citesort}
%\bibliography{spin2010proc_okada}
%\include{spin2010proc_okada.bbl}

\providecommand{\newblock}{}

\end{document}